%% file: main.tex
\documentclass[sigconf]{acmart}

\usepackage{algorithm}
\usepackage{algpseudocode}

\setcopyright{cc}
\setcctype{by}
\acmConference[RecSys '26]{The 20th ACM Conference on Recommender Systems}
  {September 28--October 2, 2026}{Minneapolis, MN, USA}
\acmYear{2026}
\copyrightyear{2026}
\acmDOI{XXXXXXX.XXXXXXX}
\acmISBN{979-8-4007-XXXX-X/2026/09}

\title{Do Sequential Recommendation Benchmarks Really Require Higher-Order Sequence Modelling?}

\author{Aleksandr V. Petrov}
\affiliation{
    \institution{Spotify}
    \city{Glasgow}
    \country{United Kingdom}
}
\email{aleksandrv@spotify.com}

\author{Praveen Chandar}
\affiliation{
    \institution{Spotify}
    \city{Washington, DC}
    \country{United States}
}
\email{praveenr@spotify.com}

\author{Paul N. Bennett}
\affiliation{
    \institution{Spotify}
    \city{Boston}
    \country{United States}
}
\email{pbennett@spotify.com}

\author{Hugues Bouchard}
\affiliation{
    \institution{Spotify}
    \city{Barcelona}
    \country{Spain}
}
\email{hb@spotify.com}

\author{Mounia Lalmas}
\affiliation{
    \institution{Spotify}
    \city{London}
    \country{United Kingdom}
}
\email{mounia@acm.org}

\begin{document}

\begin{abstract}

Sequential recommenders increasingly use language-model architectures designed to capture complex, context-dependent interactions. Yet it remains unclear whether widely used benchmarks actually require this modelling capacity.
We investigate this question using two simple, recency-weighted pairwise probes that do not learn higher-order sequence representations: Sequential Rules (SeqRules) and our Probabilistic Collaborative Transition Model (PCTM).
Using the evaluation protocol of eSASRec, at least one probe exceeds our eSASRec reproduction by 15--38\% on three Amazon datasets and by 4.4\% on MovieLens-1M, but trails it by 27.3\% on MovieLens-20M.
On the four remaining datasets, at least one probe also outperforms our sampled-softmax SASRec reproduction by 9--28\%, suggesting that these widely used benchmarks are poorly suited to measuring gains from higher-order sequence modelling.
More broadly, comparing Transformer-based models against strong recency-weighted pairwise probes provides a concrete test of whether a benchmark can meaningfully measure gains from higher-order sequence modelling.

\end{abstract}



\maketitle

\section{Do Sequential Recommendation Benchmarks Require Higher-Order Sequence Modelling?}

Recent advances in sequential recommendation have been driven by architectures originally developed for language modelling. Transformer encoders and generative retrieval models have demonstrated impressive performance by capturing complex sequential dependencies~\cite{tikhonovich2025esasrec,rajput2023tiger,hou2025actionpiece}.
However, adopting these architectures does not show that benchmark performance depends on the same complex interactions as language: it may instead be explained by item popularity, pairwise transitions, and recency. This raises a fundamental question: \emph{do current sequential recommendation benchmarks require higher-order sequence modelling?} Our claims concern what these protocols can identify, not the intrinsic complexity of deployed recommendation tasks.

Prior work highlights the importance of evaluating sophisticated models against strong baselines. A reproducibility study of non-sequential recommendation~\cite{dacrema2019progress} reproduced only 7 of 18 neural methods, six of which were often outperformed by simple heuristic baselines. These findings highlight the importance of establishing whether apparent gains truly arise from additional modelling capacity.

Similar questions have been raised for sequential recommendation. Prior work has identified issues including dataset–task mismatch, temporal leakage, and sampled negatives~\cite{hidasi2023flaws}, while other studies argue that MovieLens primarily reflects preference elicitation and recommender exposure rather than natural consumption~\cite{fan2024movielens}. The most closely related work tests whether item order carries predictive signal by shuffling histories~\cite{klenitskiy2026patterns}. Orthogonally, we ask whether that signal requires \emph{higher-order} interactions: joint, non-separable effects among history items, rather than long-range but itemwise-decomposable pairwise evidence. A benchmark may therefore be order-sensitive yet pairwise-sufficient.

Recency-based Sampling of Sequences (RSS) improves neural sequential recommenders by placing greater emphasis on recent interactions while retaining older events~\cite{petrov2022recency,petrov2025rss}, further highlighting the importance of recency in sequential recommendation. Separately, pairwise transitions have shown strong predictive performance in next-city recommendation~\cite{petrov2021nextcity}. Motivated by these observations, we ask how much predictive power can be captured by simple recency-weighted pairwise transition models, and when higher-order sequence modelling provides measurable gains beyond them.

To answer these questions, we compare increasingly expressive models ranging from last-item transition baselines to two history-aware models: Sequential Rules (SeqRules)~\cite{ludewig2018evaluation} and our \emph{Probabilistic Collaborative Transition Model (PCTM)}. Both aggregate pairwise transition information across a user's interaction history without learning higher-order sequence representations. Together, they serve as \emph{capacity probes}. 
For each dataset, we take the better of the SeqRules and PCTM scores as the \emph{pairwise envelope}. If a Transformer cannot exceed this score, the benchmark provides little evidence that higher-order sequence modelling improves prediction.

We evaluate these models using the benchmark, data splits, and evaluation protocol released with eSASRec~\cite{tikhonovich2025esasrec}, enabling direct comparison with both sampled-softmax SASRec (SAS+) and eSASRec. The reproduction matches the reported eSASRec results within 0.0033 NDCG across all ten model--dataset pairs, providing confidence in the comparison.

\section{Capacity Probes}
We evaluate transition-based models, from last-item baselines to history-aware models, while deliberately excluding higher-order sequence modelling.

\paragraph{Last-item controls}
We first establish three controls that condition only on the user's most recent interaction. 
\emph{(i) MC} is the empirical first-order Markov model, ranking candidates by empirical transition probability. 
\emph{(ii) FMC} follows the reduction analysis of SASRec~\cite{kang2018sasrec}, parameterising directional transitions through separate source and target embeddings, and is trained using SASRec's one-negative binary cross-entropy objective. Previous work has shown that replacing this objective with sampled or full softmax substantially improves performance~\cite{klenitskiy2023dross,zhai2023retrieval}. We therefore include \emph{(iii) FMC+}, which uses the same architecture trained with full-catalogue softmax cross-entropy. None of these models employs a sequence encoder.

Last-item models capture only immediate transitions. We therefore introduce two history-aware capacity probes that aggregate pairwise transition information across multiple past interactions while deliberately avoiding higher-order sequence representations.

\paragraph{Sequential Rules (SeqRules)}
We use a sparse implementation of Sequential Rules~\cite{ludewig2018evaluation} and evaluate it under the released eSASRec protocol. We tune rule distance and decay, row pruning, history length and weighting, and optional IDF weighting.

\paragraph{Probabilistic Collaborative Transition Model (PCTM)} Our model, PCTM, treats each history item as a separate source of evidence about the next item. For
each source item $b$, it estimates a directional next-item distribution $\hat P(a\mid b)$ from later-occurring items, weighting nearby occurrences more heavily. Bayesian smoothing pulls low-evidence distributions towards a uniform catalogue prior: frequent sources rely primarily on observed transitions, whereas rare sources fall back more heavily on the prior. The smoothing parameter $\tau$ controls how much evidence is required.

Recency enters at two stages. During model construction, nearby transitions contribute more to the estimated next-item distributions than distant ones. At recommendation time, predictions from recent history items receive greater weight while older interactions continue to provide lower-weight context. We consider exponential decay and a head-tail kernel motivated by RSS~\cite{petrov2022recency,petrov2025rss}, which reserves most weight for recent interactions while retaining information from the more distant history.

Each history item therefore acts as an expert over the next item. PCTM combines these experts by multiplying their conditional probabilities, implemented as a weighted sum of log-probabilities for numerical stability. A separate popularity term corrects for global popularity effects. PCTM learns no embeddings or sequence encoder. The formal model definition, history kernels, and recommendation algorithm are provided in the supplementary materials (Algorithms 1–2), while the selected hyperparameters are summarised in Table~\ref{tab:params}.

Both probes use distant events but sum itemwise evidence, never estimating a joint term such as $P(a\mid b,c)$. The stronger model defines the pairwise envelope. If Transformer-based models fail to exceed it, the benchmark provides little evidence that higher-order modelling adds predictive value; otherwise, the remaining gap measures their advantage beyond these pairwise probes.

\section{Preliminary Evidence}

We evaluate all models using the processed data splits, evaluation code, and protocol released with eSASRec~\cite{tikhonovich2025esasrec}, preserving full-catalogue ranking, seen-item filtering, and the absence of  sampled test negatives. This enables a like-for-like comparison with the reported eSASRec results. We reproduce both SAS+ and eSAS on all five datasets. All reported values are therefore our matched-protocol runs. Hyperparameters are selected on an inner training holdout before retraining each model on the full training split for evaluation on the untouched test set.

Code, frozen configurations, data-preparation scripts, and commands for reproducing Table~\ref{tab:results} are available in our public repository.\footnote{\url{https://github.com/spotify-research/sequential-capacity-probes}} The repository reconstructs all five benchmark splits from public sources, verifies their hashes, pins the upstream eSASRec implementation and software environment, and automatically checks reproduced results against the reported values.

\begin{table}[t]
  \caption{Full-catalogue NDCG@10. All values are our matched-protocol runs.}
  \label{tab:results}
  \centering
  \footnotesize
  \setlength{\tabcolsep}{2.5pt}
  \begin{tabular}{lrrrrrrr}
    \toprule
    Data & MC & FMC & FMC+ & SAS+ & eSAS & SeqRules & PCTM \\
    \midrule
    Beauty & .0492 & .0481 & .0531 & .0537 & .0524 & .0605 & \textbf{.0635} \\
    Sports & .0251 & .0244 & .0271 & .0315 & .0324 & \textbf{.0371} & .0368 \\
    Toys   & .0548 & .0554 & .0588 & .0575 & .0533 & .0730 & \textbf{.0738} \\
    ML-1M  & .1240 & .1124 & .1206 & .1662 & .1739 & .1505 & \textbf{.1815} \\
    ML-20M & .1050 & .0807 & .1034 & .1806 & \textbf{.1969} & .1115 & .1431 \\
    \bottomrule
  \end{tabular}
\end{table}

\paragraph{How far does the last item go?}
As shown in Table~\ref{tab:results}, FMC+ performs surprisingly well despite relying only on the user's most recent interaction. On the Amazon benchmarks, it is close to both Transformer baselines on Beauty, exceeds them on Toys, and reaches 84\% of eSASRec's performance on Sports. This is consistent with previous work classifying these datasets as predominantly Markov-style~\cite{klenitskiy2026patterns}. However, the pairwise envelope improves on the best last-item model by 20–46\%, showing that additional predictive signal exists beyond the final interaction. These results refine that diagnosis: although last-item transitions capture much of the available signal, incorporating several recent interactions remains beneficial.

\paragraph{Does that context require higher-order modelling?}
Table~\ref{tab:results} shows that on four of the five benchmarks, the answer is no. SeqRules and PCTM aggregate only pairwise transition information across the interaction history, yet the pairwise envelope exceeds our eSASRec reproduction by 21.3\%, 14.8\%, and 38.4\% on the Amazon benchmarks and by 4.4\% on MovieLens-1M. One explanation is that the probes supply explicit pairwise estimates and a fixed recency bias that attention must learn from sparse data. PCTM outperforms SeqRules on four of five datasets, suggesting that Bayesian smoothing and weighted log pooling provide a stronger pairwise reference, although an ablation is needed to isolate their individual contributions. More importantly, the capacity probes distinguish benchmarks where pairwise transitions explain most of the predictive signal from those where Transformer models retain an advantage beyond these probes.

\section{Conclusions}

Only on ML-20M does our eSASRec reproduction maintain a substantial advantage: the pairwise envelope reaches 0.1431 versus 0.1969, leaving 27.3\% of its NDCG unexplained. Among the benchmarks we evaluate, only ML-20M shows a clear predictive advantage beyond these pairwise probes.

Our conclusion is therefore deliberately narrow. Improvements demonstrated only on small Amazon benchmarks, or only relative to generative recommenders such as TIGER~\cite{rajput2023tiger}\footnote{Following common practice, a paper-to-paper comparison makes PCTM's NDCG@10 appear 65.4\%, 63.5\%, and 70.8\% higher than TIGER's reported values on Beauty, Sports, and Toys. However, TIGER has no official implementation, so these gains cannot be verified under identical data splits and evaluation protocol. This illustrates why paper-to-paper comparisons alone should not support state-of-the-art claims.} should not be interpreted as evidence of state-of-the-art sequence modelling without comparison against strong transition-based baselines. More broadly, we argue that capacity probes complement existing benchmark evaluation by assessing whether reported improvements reflect higher-order sequence modelling or can already be explained by pairwise transition models.

This extended abstract reports work in progress. The full study will test temporal and industrial data, learned recency kernels, and pooling under correlated histories and mode changes.

\bibliographystyle{ACM-Reference-Format}
\bibliography{references}

\input{supplement}

\end{document}

%% file: supplement.tex
\appendix
\section*{Supplementary Materials}

\paragraph{Formal PCTM definition.}
For item catalogue $\mathcal I$, let $N^{(d)}_{ba}$ count training occurrences of
candidate $a$ appearing $d$ positions after source item $b$. PCTM forms
distance-weighted causal counts
\begin{equation}
 C_{ba}=\sum_{d=1}^{W}\gamma_dN^{(d)}_{ba},
 \qquad C_{b\cdot}=\sum_i C_{bi},
 \label{eq:pctm-counts}
\end{equation}
where $\gamma_1=1$ fixes the count scale. Let $u_a=1/|\mathcal I|$ be a uniform
catalogue prior. The prior
$\boldsymbol\theta_b\sim\operatorname{Dirichlet}(\tau\mathbf u)$ gives
\begin{equation}
 \hat P(a\mid b)=\frac{C_{ba}+\tau/|\mathcal I|}{C_{b\cdot}+\tau}.
 \label{eq:pctm-conditional}
\end{equation}
Here $\tau/|\mathcal I|$ is each candidate's pseudo-count. For
$h=(h_1,\ldots,h_m)$ (newest last), let $L=\min(m,L_{\max})$, where $L_{\max}$
is the history cap, and rank by
\begin{equation}
 s_{\mathrm{PCTM}}(a\mid h)=
 \sum_{j=1}^{L}w_j\log\hat P(a\mid h_{m-j+1})
 \;+\;\lambda\log p_{\mathrm{pop}}(a).
 \label{eq:pctm-score}
\end{equation}
Equation~\eqref{eq:pctm-score} is a weighted product of per-item experts in log
space; candidate-independent denominators cancel, giving
Algorithm~\ref{alg:pctm-recommend}.

\paragraph{History kernels.}
The nonnegative history weights sum to one, with $j=1$ denoting the newest
item. Exponential weighting is
\begin{equation}
 E(r)_j=\frac{r^{j-1}}{\sum_{\ell=1}^{L}r^{\ell-1}}.
 \label{eq:exponential-kernel}
\end{equation}
For $L>k$, the head--tail kernel assigns mass $\rho$ to the $k$ newest items
and shares the remaining mass uniformly among older items:
\begin{equation}
 H(k,\rho,r)_j=
 \begin{cases}
 \displaystyle\frac{\rho r^{j-1}}{\sum_{\ell=1}^{k}r^{\ell-1}},&j\leq k,\\[3pt]
 \displaystyle\frac{1-\rho}{L-k},&j>k.
 \end{cases}
 \label{eq:head-tail-kernel}
\end{equation}
For $L\leq k$, weights are renormalised; we also test inverse decay.

\begin{table}[H]
  \caption{Selected hyperparameters. $L_{\max}$ is the history cap, $K$ row
  pruning, $I$ IDF, and $Q/D$ quadratic/div weighting.}
  \label{tab:params}
  \centering
  \tiny
  \setlength{\tabcolsep}{1.5pt}
  \resizebox{\columnwidth}{!}{
  \begin{tabular}{lccrcccrrrrrr}
    \toprule
    & \multicolumn{6}{c}{PCTM} & \multicolumn{6}{c}{SeqRules} \\
    \cmidrule(lr){2-7}\cmidrule(lr){8-13}
    Data & $\gamma_d$ & $W$ & $\tau$ & $w_j$ & $L_{\max}$ & $\lambda$
         & $W$ & $g$ & $K$ & $I$ & $L_{\max}$ & $v$ \\
    \midrule
    Beauty & $1/\sqrt d$ & 10 & $11{,}250$ & $E(.6)$ & 50 & $-.04$
           & 10 & log & 100 & N & 10 & Q \\
    Sports & $1/\sqrt d$ & 10 & $15{,}000$ & $H(10,.8,.7)$ & 50 & $0$
           & 30 & log & 0 & N & 20 & D \\
    Toys & $.7^{d-1}$ & 20 & $15{,}000$ & $H(7,.8,.6)$ & 50 & $0$
         & 30 & log & 0 & N & 10 & D \\
    ML-1M & $1/d$ & 20 & $150$ & $H(7,.8,.6)$ & 200 & $-.40$
          & 10 & div & 200 & Y & 10 & Q \\
    ML-20M & $.5^{d-1}$ & 10 & $225$ & $H(7,.8,.6)$ & 200 & $-.50$
           & 20 & div & 100 & Y & 5 & Q \\
    \bottomrule
  \end{tabular}}
\end{table}

\paragraph{Computation.}
PCTM counts transitions offline (Algorithm~\ref{alg:pctm-build}) and adds sparse
history evidence online (Algorithm~\ref{alg:pctm-recommend}). Equal-score ties
use ascending SHA-256 digest of UTF-8 external item IDs, a deterministic
evaluation convention outside PCTM.

\begin{algorithm}[H]
  \caption{\textsc{BuildPCTM}$(\mathcal D,W,\boldsymbol\gamma)$}
  \label{alg:pctm-build}
  \small
  \begin{algorithmic}[1]
    \Require Training sequences $\mathcal D$; maximum distance $W$
    \Require $\gamma_1=1$; exponential decay means $\gamma_d=q^{d-1}$
    \Ensure Sparse transition matrix $C$; popularity $p_{\mathrm{pop}}$
    \State $C\gets\{\}$; item frequencies $f\gets\mathbf 0$
    \ForAll{$x=(x_1,\ldots,x_n)\in\mathcal D$}
      \For{$t=1,\ldots,n$}
        \State $f_{x_t}\gets f_{x_t}+1$
        \For{$d=1,\ldots,\min(W,n-t)$}
          \State $C_{x_t,x_{t+d}}\gets C_{x_t,x_{t+d}}+\gamma_d$
        \EndFor
      \EndFor
    \EndFor
    \ForAll{$a\in\mathcal I$}
      \State $p_{\mathrm{pop}}(a)\gets
        \dfrac{f_a+1}{\sum_i f_i+|\mathcal I|}$
    \EndFor
    \State \Return $C,p_{\mathrm{pop}}$
  \end{algorithmic}
\end{algorithm}

\begin{algorithm}[H]
  \caption{\textsc{RecommendPCTM}$(h,C,p_{\mathrm{pop}},\tau,\mathbf w,L_{\max},\lambda,K)$}
  \label{alg:pctm-recommend}
  \small
  \begin{algorithmic}[1]
    \Require History $h=(h_1,\ldots,h_m)$; statistics $C,p_{\mathrm{pop}}$
    \Require $\tau$; $L\gets\min(m,L_{\max})$; renormalised weights $\mathbf w_{1:L}$
    \Require Popularity weight $\lambda$; return size $K$
    \Ensure $K$ protocol-eligible recommendations
    \ForAll{$a\in\mathcal I$}
      \State $s_a\gets\lambda\log p_{\mathrm{pop}}(a)$
        \Comment{keep all items; this is 0 if $\lambda=0$}
    \EndFor
    \For{$j=1,\ldots,L$}
      \State $b\gets h_{m-j+1}$ \Comment{$j=1$ is the newest item}
      \ForAll{$a$ with $C_{ba}>0$}
        \State $s_a\gets s_a+w_j\log\!\left(1+\frac{|\mathcal I|C_{ba}}{\tau}\right)$
      \EndFor
    \EndFor
    \ForAll{$a\in h$}
      \State $s_a\gets-\infty$ \Comment{if repeated targets are excluded}
    \EndFor
    \State \Return first $K$ eligible items by decreasing $s_a$
  \end{algorithmic}
\end{algorithm}